\documentclass[12pt]{article}
\usepackage{amsmath}
\usepackage{amssymb}
\usepackage{epsfig}
\usepackage{cite}
\usepackage{color,colordvi}

\begin{document}
\begin{flushright}
\vskip-0.8cm
{\small  MPP--2015--215},
{\small LMU--ASC 59/15}
\end{flushright}
\vspace{0.01cm}
\vskip-0.5cm
\begin{center}
{\Large\bf   
Classical Limit of Black Hole Quantum $N$-Portrait and BMS Symmetry}

\end{center}

\vspace{-0.1cm}

\begin{center}

{\bf Gia Dvali}$^{a,b,c}$,  {\bf Cesar Gomez}$^{b,e}$\footnote{cesar.gomez@uam.es} and {\bf Dieter L\"ust}$^{a,b}$\footnote{dieter.luest@lmu.de}

\vspace{.6truecm}


{\em $^a$Arnold Sommerfeld Center for Theoretical Physics\\
Department f\"ur Physik, Ludwig-Maximilians-Universit\"at M\"unchen\\
Theresienstr.~37, 80333 M\"unchen, Germany}


{\em $^b$Max-Planck-Institut f\"ur Physik\\
F\"ohringer Ring 6, 80805 M\"unchen, Germany}

{\em $^c$Center for Cosmology and Particle Physics\\
Department of Physics, New York University, \\ 
4 Washington Place, New York, NY 10003, USA}

{\em $^e$
Instituto de F\'{\i}sica Te\'orica UAM-CSIC\\
Universidad Aut\'onoma de Madrid,
Cantoblanco, 28049 Madrid, Spain}\\

\end{center}

\vspace{0.1cm}

\begin{abstract}
\noindent  
 {
\small
Black hole entropy,  denoted by $N$,  in (semi)classical limit is infinite.  This scaling 
reveals a very important information about the qubit degrees of freedom that carry black hole entropy.   
Namely, the multiplicity of qubits scales as $N$, whereas their energy gap and their coupling as $1/N$.  
Such a behavior is indeed exhibited by Bogoliubov-Goldstone degrees of freedom of a quantum-critical  
state of $N$ soft gravitons  (a condensate or a coherent state) describing the black hole quantum portrait. They can be viewed as the Goldstone modes of a broken symmetry acting on the graviton condensate.  In this picture Minkowski space 
naturally emerges as a coherent state of $N=\infty$ gravitons of {\it infinite} wavelength and it carries an infinite entropy. In this paper we ask what is the geometric meaning (if any) of the classical limit of this symmetry. 
 We argue that the infinite-$N$ limit of Bogoliubov-Goldstone modes of critical graviton condensate is described by recently-discussed classical BMS super-translations broken by the black hole geometry. 
 However, the full black hole information can only be recovered for finite $N$, since the recovery time becomes infinite 
 in classical limit in which $N$ is infinite.}

\end{abstract}

\thispagestyle{empty}
\clearpage

\section{Entropy Scaling and Information Qubits}

  Black hole entropy scales as area in Planck units \cite{Bek}, 
    \begin{equation}
 N \, =\, {R^2\over L_P^2} \, .
 \label{N}
 \end{equation}
  Here $R = G_NM$ is the gravitational radius of black hole of mass $M$ and $G_N$ is the Newton's constant. 
  $L_P$ is Planck length, which in terms of $G_N$ and Planck's constant $\hbar$ is defined as 
     \begin{equation}
 L_P^2 \, \equiv \, \hbar G_N \, .
 \label{LP}
 \end{equation}
  We have set the speed of light and all the other numerical factors to one.  
  
    The Bekenstein's formula tells us that the black hole is a quantum state with degeneracy of micro-states 
scaling exponentially with $N$.   The crucial question is:  What are the microscopic  
qubit degrees of freedom describing the above degeneracy? 

 In answering this question, first notice that some important information about these qubits can be gained from analyzing the classical and semi-classical  limits of the Bekenstein's formula  and matching it with well-established known properties of classical  black holes \cite{usBH1, usBH2, gold2, gold3}. 
 
 The limits of interest are the ones in which the quantum back-reaction on classical geometry vanishes.   
  We shall distinguish the two limits of this sort.  
  
   The first we shall refer to as the {\it semi-classical limit}. 
  In this limit, we keep both $\hbar$ and $R$ finite, since we want to keep notions of geometry as well as of quantum 
  mechanics well-defined.  The only consistent way of achieving this is to take 
  $M \rightarrow \infty$, $G_N \rightarrow 0$, but in such a way that we keep their product ($R$) finite. 
   Thus, with finite $\hbar$, in order to keep geometry exact, we must take the black hole to be {\it infinitely massive}, 
 and simultaneously, gravity to be {\it infinitely weak}.  
  
   The second limit, is the classical one, $\hbar \rightarrow 0$. In this limit, we can keep the geometry exact, even for finite values of $M$ and $G_N$. 
   
    We now observe that no matter which of the two limits we choose, we have $L_P \rightarrow 0$ and thus, 
    the black hole entropy becomes infinite! 
   
   Thus, both in classical and semiclassical limits black hole carries an infinite amount of information. 
  
   The reconciliation of the above scaling with the no-hair properties of classical black holes, reveals a very valuable information about the entropy-carrier degrees of freedom. Namely \cite{usBH2, gold1, gold2, gold3}:    
   
  $~~~$
  
  {\it  1) Number of qubits scales as $N$};  
  
  $~~~$
   
 {\it 2) Their energy gap and their coupling scales as $1/N$}.
  
  $~~~$
  
  Putting it simply, although in classical limit the number of qubits becomes infinite, they decouple and resolving the degeneracy of micro-states becomes infinitely-hard. Correspondingly, the no-hair property is valid for any finite observation time of a classical black hole.  
 
   The composite multi-graviton portrait of a black hole \cite{usBH1,usBH2} provides a natural candidate for information-carrier qubit degrees 
   of freedom in form of the   Bogoliubov-Goldstone degrees of freedom of the soft-graviton condensate.  
   According to this picture, the black hole is well-described as a bound-state of $N$ soft gravitons. 
   Examining the coupling of these gravitons, it is evident that the condensate is at a quantum critical point. 
    The qubit degrees of freedom are then identified with $N$ Bogoliubov degrees of freedom that populate the spectrum within  $1/N$ energy gap.  For large $N$, their coupling scales as $1/N$. 
     
    These degrees of freedom can be described as Goldstone modes of certain rank-$N$ symmetry group acting on the graviton condensate\cite{gold1,gold2, gold3} \footnote{ Here we use the word rank to refer to the number of generators of the algebra that creates Goldstone modes. At this point we do not need to commit ourselves to any particular  realization of this symmetry.}.  It has been appreciated that Goldstone interpretation of these modes, 
 with the number of broken generators  as well as the spontaneous breaking order parameter scaling as $N$, 
 automatically accounts for the right $1/N$ scaling behavior both of the energy gap as well as of the coupling of Goldstone-qubits. Correspondingly, this picture naturally accounts for various time-scales of information-processing, such as, generating chaos and scrambling information within the time $t_{scr} \sim \sqrt{N} L_Pln(N)$ \cite{scrambling}, as well as, generating large entanglement
at the quantum critical point \cite{nico,gold1,gold2,gold3}.   The information-recovery time scales as $t_{inf} \sim N^{3/2}L_P$ and becomes infinite both in  classical and semi-classical limits.    

  Thus, in this picture, a classical black hole carries infinite amount of hair, but it is undetectable, because the resolution time is infinite.  For finite $N$, the hair carries all the information about the black hole quantum state, including 
 some global  charges that it may swallow \cite{baryon}.  The same hair manifests itself in 
  $1/N$-corrections to Hawking radiation\cite{usBH1, usBH2}, which give deviations from thermality and encode the information. Black hole hair also manifests itself in $1/N$-corrections to 
 the probe particle scattering at the critical graviton condensate \cite{ds}. These corrections measure the $1/N$ quantum deviation  from the classical metric motion. The analogous $1/N$-corrections corrections measure deviation from
 entropy suppression, $e^{-N}$, in black hole formation in scattering of two energetic gravitons into the $N$ soft ones\cite{scattering}.  

   Notice, that  for finite $N$ the symmetry need not be exact, and Goldstones are not exactly gapless,  
 but the energy gap must vanish for infinite $N$.    
     Below, in order to keep our discussion maximally general, we shall simply refer to this symmetry as rank-$N$ group. 
      
     The question we would like to ask in the present paper, is:   What is the geometric interpretation, if any, of this 
  symmetry, with spontaneously-broken $N$ generators,  
   in the classical limit? 
    We shall try to argue that the answer is the classical BMS symmetry \cite{BMS}, recently studied   
  by Strominger  et al \cite{stro1,stro2,stro3}.    
   
  In order  to answer this question, we shall first make the connection to the limiting case of Minkowski space, describing it as a coherent state of gravitons. 
    
     \section{Minkowski Vacuum as Coherent state}   
   
  \subsection{Shift of Scalar VEV as Coherent States}     
 
 We shall start with the toy model of pure scalar gravity first. 
  Consider a massless scalar field  $\phi(x_{\mu})$ in $3$-space dimensions.  In order to regularize the problem, 
 let us introduce a small mass $m$ and then take the zero mass limit.  The Lagrangian is,  
  \begin{equation}
   {\mathcal L} \, = \,  { 1\over 2} \left ( (\partial_{\mu}\phi)^2  \, - \, m^2\phi^2  \right)\, . 
 \label{L1}
 \end{equation}
 In $m=0$ zero limit,  we have a continuous symmetry $\phi \rightarrow \phi + v$, where $v$ is an arbitrary constant. 
 The  Noether current of this symmetry is $J_{\mu} \equiv \partial_{\mu} \phi$, which for 
 $m=0$ is conserved by the equation of motion. For nonzero mass, the divergence of the current 
 is proportional to the mass term, which breaks the shift symmetry explicitly
   $\partial_{\mu}J^{\mu} \, = \, -m^2\phi$.  We shall work in the finite volume  $V = (2\pi R)^3$ and then take the
   infinite volume limit.   The standard mode-expansion of the field $\phi$ is,    
     \begin{equation}
  \phi (x) \, = \,  \int  \, {d^3 (kR) \over \sqrt{V \omega_k}}  \, ( e^{-i(\omega_kt - \vec{k}\vec{x} )}  a_{\vec{k}} \, + \, 
  e^{i(\omega_kt - \vec{k}\vec{x} )}  a_{\vec{k}}^+) \,.   
 \label{Fieldoperator}
 \end{equation}  
 with $\omega_k = \sqrt{m^2 + |k|^2}$. 
  The charge operator is, 
      \begin{equation}
 Q \equiv \int d^3x J_0 \, = \int d^3x \partial_t \phi (x) \, = \, -i q  ( e^{-imt}  a_{0} \, - 
 e^{imt}  a_{0}^+) \, 
 \label{Shiftgeneral}
 \end{equation}  
  where $q \equiv \sqrt{Vm}$.  Notice,  in this parameterization $Q$ has a dimensionality of  $\phi^{-1}$.  This is because the shift parameter $v$ has the same  dimensionality as $\phi$. Now acting with the shift operator on the vacuum state $|0\rangle$, we create another state $|v\rangle$, 
       \begin{equation}
 |v\rangle = e^{-ivQ} |0\rangle \, .  
 \label{Shiftvacuum}
 \end{equation}    
 Notice that from the form of the shift generator (\ref{Shiftgeneral}), it is clear that the state $|v\rangle$ is  a coherent state of zero momentum quanta, characterized by the complex parameter $\sqrt{N_0}e^{i\theta}$, where the modulus measures the occupation number  $N_0 = (qv)^2$ and the phase  is $\theta = mt$. It can be written in the form, 
 \begin{equation}
 	\label{cohvac}
	 | v\rangle \,  =  e^{-\frac{N_0}{2}} \sum_{n_0=0}^{\infty}  \frac{(N_0e^{i2mt})^{\frac{n_0}{2}}}{\sqrt{n_0 !}} | n_0 \rangle,
\end{equation}	
 where $|n_0\rangle$ are Fock states of zero momentum quanta of occupation number $n_0$. 
 
 Notice, that from the scalar product of two coherent states 
  \begin{equation}
 | \langle v |v'  \rangle|^2 \, = \, {\rm e}^{- q^2 (v - v')^2} \,,  
 \label{projection}
 \end{equation} 
 it is clear that, for finite $v,v'$, they become orthogonal only for $q \rightarrow \infty$ limit. 
 If in this limit we keep $m$ finite, the resulting state will describe a coherently oscillating scalar field 
 of amplitude $v$ and frequency $m$. The expectation value of the scalar field over such a coherent state 
 behaves as a classical field,  
   \begin{equation}
  \langle v|\phi |v\rangle \, = \, v e^{-imt} \, + {\rm h.c.} \, .
 \label{VEVm}
 \end{equation} 
 
  However, if simultaneously, we take the limit, 
 $m \rightarrow 0$, the states with different values of $v$ become degenerate vacua. 
  The expectation value of the field $\phi$  in such a vacuum is 
  \begin{equation}
  \langle v|\phi |v\rangle \, = \, v \, .
 \label{VEV}
 \end{equation} 
 
 Thus, we see that the vacua with constant values of the scalar field can be understood as coherent states 
 with infinite occupation number of infinite wave-length quanta.

 \subsection{Gravitational Minkowski Vacua as Coherent States} 
 
  We are now ready to apply the coherent state picture to Minkowski space in theory with gravity. 
 Since for Minkowski space all the curvature invariants vanish and we are dealing with infinitely soft gravitons, 
 we shall work within the weak field expansion. Consider a linearized sourceless equation for spin-2 field,  
 \begin{equation}
  {\mathcal E} h_{\mu\nu} \, - m^2 \, (h_{\mu\nu} - \eta_{\mu\nu} h)  \, =   \, 0 \,,  
 \label{cc}
 \end{equation} 
  where ${\mathcal E} h_{\mu\nu} \, \equiv \,  \Box h_{\mu\nu} \, - \, \eta_{\mu\nu} \Box h \, - \, \partial_{\mu} \partial^{\alpha} h_{\alpha\nu} \, - \, \partial_{\nu} \partial^{\alpha} h_{\alpha\mu}  \, - \, \partial_{\mu} \partial_{\nu} h \, - \, 
  \eta_{\mu\nu} \partial^{\alpha}\partial^{\beta} h_{\alpha\beta}$
 is the linearized Einstein tensor and $h\equiv h_{\alpha}^{\alpha}$. 
     As we did in the scalar case, we have regularized the system by adding a small 
  Pauli-Fierz mass term, which we shall later take to zero. 
  As it is well known, such a regularization introduces the $3$ additional propagating degrees of freedom in form of two 
  helicity-$1$ and one helicity-$0$ polarizations. 
  
   In the basis in which kinetik terms  are diagonal the decomposition of massive graviton  $h_{\mu\nu}$ 
 in terms of  Einsteinian helicity-$2$ tensor component $\widetilde{h}_{\mu\nu}$, 
 helicity-$1$ vector $A_{\mu}$  and a  helicity-$0$ scalar $\phi$ has the 
 following form,  
\begin{equation}
h_{\mu\nu} = \widetilde{h}_{\mu\nu} +  \partial_{\mu} A_{\nu}  +  \partial_{\mu} A_{\nu}  + 
\frac{1}{6} \eta_{\mu\nu}\phi + \frac{1}{3} \frac{\partial_\mu \partial_\nu}{m^2} \phi. 
\label{helicities} 
\end{equation}
The massive graviton $h_{\mu\nu}$ is invariant under the gauge shift
$A_{\mu} \rightarrow A_{\mu} - \xi_{\mu}, ~~  \tilde{h}_{\mu\nu} \rightarrow \tilde{h}_{\mu\nu} + \partial_{\mu} \xi_{\nu}
+ \partial_{\nu} \xi_{\mu}$, which acts on the Einsteinian component  $\tilde{h}_{\mu\nu}$ as the usual linear coordinate transformation.     
  Below, we shall set the  helicity-$1$ component to zero.

 For $m=0$ there exists a conserved current: $J_{\gamma (\mu\nu)} = 
 \partial_{\gamma} h_{\mu\nu} - \eta_{\mu\nu} \partial_{\gamma} h - 
 (\partial_{\mu}h_{\gamma\nu}  +  \partial_{\nu}h_{\gamma\mu} ) + {1\over 2} (\eta_{\gamma \mu} \partial_{\nu} + \eta_{\gamma\nu}\partial_{\mu} ) h
 + \eta_{\mu\nu} \partial^{\beta} h_{\gamma\beta}$.   
  The divergence of this current 
 is simply the linearized Einstein tensor, $\partial^{\gamma} J_{\gamma (\mu\nu)} = {\mathcal E} h_{\mu\nu}$,  which vanishes by the equation of motion of the massless theory. 
 
 Moreover,  notice that the trace of the current over $\mu,\nu$ indexes, 
 \begin{equation} 
 J_{\gamma} \, \equiv \, \eta^{\mu\nu} J_{\gamma (\mu\nu)} \, = \, 2(\partial^{\nu} h_{\gamma \nu} - \partial_{\gamma} h ) \, , 
 \label{trace}
 \end{equation} 
  vanishes by the Pauli-Fierz constraint, which can be obtained by taking the divergence of the  equation of motion
  (\ref{cc}).
   We shall use this fact to make a shortcut for interpreting Minkowski as a coherent state. 
  The current (\ref{trace}) vanishes on the Pauli-Fierz massive graviton $h_{\mu\nu}$,  but it is non-zero on 
  the Einsteinian $\tilde{h}_{\mu\nu}$  and helicity-zero $\phi$ components separately.  We are not interested in the total Pauli-Fierz graviton, but only in its Einsteinian  component $\tilde{h}_{\mu\nu}$, which decouples from 
  the rest in the massless limit.  Hence, knowing that the action of the current on $\tilde{h}_{\mu\nu}$  and  $\phi$ are 
  exactly opposite,  we can read off the shift of massless graviton from the shift of $\phi$. 
   
    From the form of the charge,  
         \begin{equation}
 Q  \equiv  - \, \int d^3x J_0 \, = 
  \, \int d^3x  2 \left ( \partial_t \, \tilde{h} - \partial^{\nu} \tilde{h}_{t\nu} \right ) \, + \, \int d^3x \partial_t \phi \, , 
 \label{Shiftgraviton}
 \end{equation}   
  it is clear that its action on $\phi$ is identical to   
  (\ref{Shiftgeneral}). Thus the action on $\tilde{h}_{\mu\nu}$ is exactly the opposite, because 
  $Q$ must annihilate on any physical state of $h_{\mu\nu}$. 
  
    We can now proceed exactly as in the case of a pure scalar field.   We act on the vacuum with the shift operator, 
        \begin{equation}
 |v\rangle = e^{-iv Q} |0\rangle \,,   
 \label{Shiftvacuum1}
 \end{equation}   
 and take the limit $m\rightarrow 0$.   The difference from the pure scalar case is that we now create the coherent states containing 
the infinite occupation numbers of both $\phi$
 and $\tilde{h}_{\mu\nu}$ quanta of infinite wave-length.   The  expectation values of the quantum operators $\hat{\tilde{h}}_{\mu\nu}$ and 
 $\hat{\phi}$ on the state $|v\rangle$ satisfies,   
 \begin{equation}
\langle v| \hat{\tilde{h}}_{\mu\nu} |v\rangle \,  = \, - {1 \over 6} \eta_{\mu\nu}  \langle v| \hat{\phi} |v\rangle \, = \, -
{1 \over 6} \eta_{\mu\nu} v\, . 
  \label{vevs}
 \end{equation}   
In this way, the expectation value of the full would-be massive graviton vanishes. However, in the  massless 
limit, the independent physical degrees of freedom  are $\phi$ and $\tilde{h}_{\mu\nu}$ and we can focus
exclusively on $\tilde{h}_{\mu\nu}$.  For the
probe sources coupled to  $\tilde{h}_{\mu\nu}$ the coherent state  $|v\rangle$ is the exact  Minkowski state. 
As it is clear from (\ref{vevs}),  the expectation value of Einsteinian graviton over this state is nothing but a classical Minkowski space metric.   
 
  Thus, we resolve Minkowski space as a coherent state of infinite occupation number of infinite wave-length 
gravitons.  Clearly, this state exhibits an infinite degeneracy with respect to the change of this number.

 \subsection{Minkowski as Coherent State and Holography}
 
 There exist a nice holographic interpretation of the previous exercise. Indeed for finite volume $V=R^3$ and finite value of the mass $m$ the energy of the coherent state $|v\rangle$ is $E= m N_0 = m^2R^3v^2$. Since this energy is localized in a volume of size $R$ we can use Bekenstein bound \cite{Bek2} on the entropy as $ER$ in $\hbar$ units. If now we identify this entropy with $N_0$ and we try to saturate the bound for the coherent state $|v\rangle$ we get the following  relation between $m$ and $R$,
 \begin{equation}
 mR \, = \, 1\, .
 \label{mrrelation} 
 \end{equation}
 In other words, in this case the coherent state $|v\rangle$ saturates the Bekenstein bound. In the limit $m=0$ we recover the infinite-volume limit and the corresponding coherent state with infinite $N_0$. However, this double-scaling  limit $m=0$ and $R=\infty$ with $mR=1$ leads to an infinite total energy $O(Rv^2)$ that vanishes only for $v=0$.
 This is very natural, since the double-scaling limit corresponds to the interpretation of the Minkowski space  
as of an infinitely-massive black hole. Although the total mass is infinite, all the curvature invariants locally vanish. 
In this case, the holographic nature of Minkowski space is simply a memory about the holographic nature of the
infinitely-massive black hole that it represents.  
 
  In summary,  we can think about the vacuum $|v\rangle$ as the double-scaling limit of the finite-volume coherent state saturating Bekenstein bound. This makes the Minkowski vacuum essentially holographic, which matches the fact that 
  it is precisely in this limit that Minkowski can be identified with an infinite mass black hole.

 The above picture of Minkowski space also  emerges as a limit from the representation of de Sitter space as of coherent state of gravitons of frequencies given by the Hubble parameter \cite{ds}. There it is shown that this description  
 fixes $H= m$, where $H$ is the Hubble parameter (i.e., cosmological constant in Planck units). 
  This implies that the  Bekenstein bound is saturated for the Hubble volume for arbitrarily-small value of 
  $H$.   
  The zero cosmological constant limit is $m=0$,  leading to Minkowski as a coherent state of infinitely-soft gravitons.  In any case, the key point to be stressed is that the coherent state representation of Minkowski space naturally  captures the  basic ingredient of holography.

   \section{Large-$N$ limit of Bogoliubov-Goldstone Modes and BMS  Symmetry.}
  
   The fact that we can understand Miknowski space as infinite occupation number coherent state of zero momentum gravitons allows us to partially address the question about the classical limit of rank-$N$ algebra. 
  Indeed, since Minkowski can be viewed as a metric seen by an observer  of an  infinitely-massive black hole, it is natural to identify the large-$N$ limit of rank-$N$ symmetry of the quantum portrait as 
 a Bondi, Metzner and Sachs (BMS)  
 symmetry \cite{BMS} of Minkowski vacuum in the spirit of \cite{stro1,stro2,stro3}.  
 With this identification, the  Bogoliubov-Goldstone modes that store black hole information, in $N \rightarrow \infty$ limit   
 are mapped on infinitely-soft graviton modes created by BMS supertranslations.   
 
   However, the story is different for the classical limit of a finite mass black hole.  As discussed above, 
   despite the fact that $N$ becomes infinite, the radius $R$ stays finite.  Therefore, the corresponding Goldstone modes must carry  the wave-length $\sim R$. Thus, they {\it cannot}  come purely from  BMS degeneracy of the asymptotic Minkowski 
 vacuum.  Thus, classical limit of the symmetry group, must be spontaneously broken by finite $R$ effects and not by the asymptotic space \cite{Hawking}.    
    The fact that these Goldstones have wavelength $R$, automatically fixes their number for any finite $\hbar$ to the value  $N$, as it is predicted by the quantum portrait.   Thus, it is natural to identify the Goldstone modes of the BMS group 
    with the  $\hbar \rightarrow 0$ limit of Bogoliubov-Goldstone modes of the critical graviton condensate.  
Below, we shall discuss this connection  in more details.    

In the last two years a new physical understanding of the BMS symmetry in asymptotically flat spaces has been put forward by Strominger and collaborators in a series of papers \cite{stro1,stro2,stro3}. The final goal of this research is among other things to generalize the ideas of holography to Minkowski or de Sitter spaces in the sense of identifying potential candidates for some "boundary" like holographic degrees of freedom.

For asymptotically flat spaces we can focus on the physics in the null infinity ${\cal I}^{\pm}$ . Its topology is $R\otimes S^2$ with $R$ parametrized by advanced and retarded time coordinate $u$ and $v$ respectively. The $S^2$ can be visualized as the section of a light cone with the null infinity. We can parametrize this $S^2$ with spherical coordinates $\theta$ and $\phi$. 

The BMS  supertranslations on ${\cal I}^+$ are simply the transformations
\begin{equation}
(u,\theta,\phi) \rightarrow (u+ f(\theta,\phi), \theta, \phi)\, ,
\end{equation}
with $f$ an arbitrary real function. The super translation is simply a translation in the retarded time dependent on the angular direction in $S^2$. The previous definition implies that we have as many generators of supertranslations as spherical harmonics,  so we can formally characterize the supertranslations as $T_{l,m}$. One of the most interesting outputs of the recent research on these asymptotic symmetries has been to realize that the Minkowski vacuum
$|0\rangle$ breaks spontaneously the invariance under supertranslations.   Or in other words, the states obtained by acting on the vacuum
\begin{equation}\label{one}
T_{l,m}|0\rangle = |l,m\rangle \, ,
\end{equation}
which are soft gravitons of infinite wave length, can be interpreted as Goldstone bosons \cite{stro1,stro3}. 

 Interestingly,  this picture resonates with the former discussion on Minkowski as a coherent state of infinite wave-length gravitons. In fact, as already pointed out, this coherent state model saturates the Bekenstein bound for a formal entropy defined as the average number of quanta $N$ of the corresponding coherent state. On the other hand, this is precisely  the scaling regime in which Minkowski coherent state represents  the infinite mass limit of a black hole.  
 Then, it becomes clear, why the entropy of the corresponding coherent state must be infinite in this regime.  
  We could think of this infinite number as being associated with the infinite number of BMS Goldstone bosons. 

The canonical example of holography has been always the black hole itself.  So, if the previous picture captures the essence of the holographic degrees of freedom,  we must look for an analog of the BMS super translations acting on the black hole horizon. This is indeed what was announced in \cite{Hawking}.  In these conditions we could think of the analog of (\ref{one}), namely
\begin{equation}
T(l,m)|BH\rangle =|l,m;BH\rangle
\end{equation}
for $|l,m;BH\rangle$ the black hole Goldstone bosons. Since in this case we are dealing with the horizon, it is natural to think of these Goldstone bosons as having wave length equal to the black hole radius, $R$. Moreover, since 
for finite $\hbar$ the black hole entropy is finite, we need somehow to reduce the effective number of these Goldstone bosons to be exactly the black hole entropy, $N$.  

At first sight the most interesting physics question could be to unveil why these finite wave-length graviton modes are Goldstone bosons relative to the black hole quantum state. If we assume that they are, then they can provide a family of inequivalent black hole quantum states nearly-degenerate in energy, and therefore, the right candidates to define the black hole entropy \footnote{The connection of BMS and the information paradox was already pointed out in \cite{stro2}.}.

In what follows, we shall like to make contact between this geometric approach to the black hole entropy and some of the main features of the portrait model of the black hole\cite{usBH1,usBH2}. In the black hole portrait the key ingredient is to identify the black hole with a self sustained condensate of gravitons at a quantum critical point. In other words, we model the black hole as a many body system of gravitons, {\it but} at a very special point, where a quantum phase transition takes place. What makes the black hole spacial, is this quantum criticality that manifest itself precisely as the appearance of nearly-gapless $N$ Bogoliubov-Goldstone modes. These Goldstone bosons are not there when we are away from criticality,  as it could be the case for a star or some other massive object. They only appear when we reach the criticality associated with the black hole formation \footnote{It is worth to notice that already in some simple toy models of critical Bose-Einstein condensates we can assign $l,m$ standard quantum numbers to the Goldstone modes and track how these many-body states become gapless at the critical point in $N\rightarrow \infty$ limit \cite{gold2}. This illustrates the crucial role of quantum criticality in generating gapless Goldstone modes.}.

In essence this microscopic picture is extremely simple and points out that once we turn on gravity and we track the gravitational self energy in quantum mechanical terms, the quantum state of the system breaks spontaneously some rank-$N$ symmetry when it becomes a black hole, i.e., when the graviton condensate is at the critical point.  
  
 This spontaneous symmetry breakdown comes with a set of Goldstone bosons that likely satisfy all the conditions to be the microscopic quantum version of the black hole super translation Goldstone bosons. In this sense, the black hole horizon as a geometric notion is simply reflecting the criticality of the many body system of gravitons, once we create a correspondence between the many-body Goldstone modes and the special BMS symmetries of the horizon.

 In summary, once we decide to promote Goldstone to be the key word underlying black hole entropy we must look for the microscopic meaning of the underlying symmetry breakdown. To our view, the criticality of the black hole portrait naturally fits the goal. 
 
 We would like to stress that there are effects, captured by the quantum $N$-portrait,  that are expected not to be visible in the classical BMS picture and their account requires the finite $N$ resolution.  For example, for finite $N$ the symmetry need not be exact and can (and in general will) be broken explicitly 
 by $1/N$-effects. This  explicit breaking gives a finite energy gap to Goldstone modes promoting them into 
 pseudo-Goldstones.  This is fully compatible with the entropy scaling, since the (pseudo)Goldstone modes do not have to be exactly gapless, but rather populate the energy gap $\sim 1/N$. 
 Moreover,  precisely these finite $N$ effects allow the recovery of black hole information within the finite time. 
 The information processing time-scales, such as the scrambling time for the critical condensate ($\sim ln(N)$) \cite{scrambling},  scale with $N$. 
 Therefore, understanding the finite $N$ portrait is crucial for the information-recovery process.  
 
  In order to make the closed contact between the large-$N$ portrait \cite{usBH1, usBH2}  and \cite{stro1, stro2, stro3}  
 one natural way would be to consider scattering of an external soft graviton at the critical $N$-graviton condensate
 and then take the semiclassical limit. 
 Moreover,  the BMS group can manifest itself also in 
black hole formation in scattering of two energetic gravitons into the $N$ soft  ones, for which the right entropy suppression appears exactly when the  $N$ gravitons are at the quantum critical point \cite{scattering}.  
 At this point, the momenta of these gravitons are $R^{-1}$, with $R$ being equal to the gravitational  
 radius corresponding to the center of mass energy $\sqrt{s}$ of the process.
 
 Since, as we have argued, the Minkowski space corresponds to an infinitely heavy  black hole, there is at least one  well defined limit, namely $N = \infty, R=\infty$, for which the scattering amplitude of \cite{scattering} can be mapped to the 
infinitely-soft graviton processes of the type \cite{stro1, stro2, stro3}  controlled by BMS supertranslations of asymptotic Minkowski space. 

   The interesting regimes to be explored are with finite $R$,  and with $N$ either finite or infinite. 
   In both cases, new types of corrections appear, both due to finite softness of gravitons and due to finite $N$. 
  One possible way to take them into the account, could be via new kind of soft theorems, which as shown recently \cite{Cachazo:2014fwa}, can arise from sub-leading contributions due to finite softness. 
  
    It is also tempting to conjecture that for finite $R$, at least to leading order in $1/N$, the scattering 
  process can be understood in terms of BMS-type Goldstone generators that are broken by finite-$R$  geometry, as opposed to the asymptotic flat space.   Then these would be the natural candidates to be identified with Bogoliubov-Goldstone modes 
  of large-$N$ graviton condensate.    
  We leave this for the future work. 
  
  The important point however is that finite $N$ effects are crucial for the information recovery and must be taken into account.

\section*{Acknowledgements}
 
 We like to thank I. Bakas and R. Isermann for useful discussions.
The work of G.D. was supported by Humboldt Foundation under Alexander von Humboldt Professorship,  by European Commission  under ERC Advanced Grant 339169 ``Selfcompletion'' and  by TRR 33 "The Dark
Universe".
The work of C.G. was supported in part by Humboldt Foundation and by Grants: FPA 2009-07908, CPAN (CSD2007-00042) and by the ERC Advanced Grant 339169 ``Selfcompletion'' . The work of D.L. was supported by  the ERC Advanced Grant 32004 ``Strings and Gravity" and also by TRR 33.

\end{document}